\documentclass[aps,pre,nofootinbib,showpacs,twocolumn]{revtex4}
\usepackage{amssymb,latexsym,mathrsfs}
\usepackage{amsfonts}
\usepackage{amsmath}
\usepackage{graphicx}

\newcommand{\beq}{\begin{equation}}
\newcommand{\eeq}{\end{equation}}
\newcommand{\beqn}{\begin{eqnarray}}
\newcommand{\eeqn}{\end{eqnarray}}
\newcommand{\bearr}{\begin{array}}
\newcommand{\enarr}{\end{array}}

\newcommand{\ket}[1]{|#1\rangle}
\newcommand{\bra}[1]{\langle#1|}
\newcommand{\ra}{\rangle}
\newcommand{\la}{\langle}

\def\bea{\begin{eqnarray}}
\def\eea{\end{eqnarray}}
\def\ba{\begin{array}}
\def\ea{\end{array}}
\def\n{\nonumber}

\begin{document}
\title{TASEP on a ring with internal degrees of freedom}
\author{ Urna Basu}
\email[E-mail address: ]{urna.basu@saha.ac.in}
\author{P. K. Mohanty}
\email[E-mail address: ]{pk.mohanty@saha.ac.in}
\affiliation{Theoretical Condensed Matter Physics Division, Saha Institute of Nuclear Physics,\\
1/AF Bidhan Nagar, Kolkata, 700064 India.}
\date{\today}
\vskip 2.cm

\begin{abstract}
A totally asymmetric exclusion process  on a ring with $\nu$ non-conserved internal degrees of freedom,
where particles hop forward with a rate  that depends on their internal state, has been studied.  We show, 
using a mapping of the model to a zero range process with $\nu$  different kinds of boxes, 
that  steady state weights  can be written in a matrix product form and calculate the spatial 
correlations exactly. A comparison of the model with  an equivalent 
conserved system reveals that unequal hopping rates  of particles belonging to different internal 
states is responsible for  the non-trivial correlations.
\end{abstract}

\maketitle

\section{Introduction}
Driven diffusive systems have been studied extensively  in recent past for their unorthodox 
non-equilibrium properties\cite{dds}.  Totally asymmetric simple exclusion process (TASEP)  
is one such system, initially introduced \cite{tasep} as a basic model for transport\cite{transport},  which 
has found applications in wide areas of physics and biology\cite{tasepRev}.  The steady state 
weights of TASEP,  which has been calculated exactly \cite{Derrida,mpaderrida,MPA}, show  collective behaviour
like boundary driven phase transitions, non-trivial density and shock-profiles. Several variations
of  this exclusion model including non-conservation of particles, multiple  species \cite{mult1,mult2} 
have been studied in different contexts. %one  can be  solved  exactly  using MPA \cite{MPA}. 

Recently  TASEP with non-conservation of particles having internal degrees of freedom has also been  
studied\cite{TASEPint}. In these models particles are allowed to enter  or exit the system from both boundaries 
and  their hop rate in the bulk  depend on the internal degree they possess.  
The explicit conditions on the rates  for  the steady state  of  a  parity-time invariant system to 
have a factorized form has been  derived. Analytical results for the 
steady state of these models for generic rates (where non-trivial spatial correlations  are expected) are not known.

In this article we introduce a totally asymmetric  exclusion process on a ring with non conserved 
internal degrees of freedom. In this model, apart from the usual forward hopping,
each particle in one  of the $\nu$ possible internal states can change 
to  any other $\nu-1$ states with different rates.  
The steady state of the model could be calculated exactly.  Unlike  TASEP on  a ring, 
which has  an uncorrelated steady state, this model  show non-zero spatial correlations 
when hop rate of  particles  depend on their internal states.

\section{The Model} 

    The model is defined on a  one dimensional lattice  with periodic boundary conditions. 
The sites, labelled by $i=1,2\dots L$, can  either be empty or  occupied  by  at 
most one  particle. Each particle  can be in one of the  $\nu$ possible internal states 
$I=1,2,....\nu$ . Correspondingly, at  the $i^{th}$ site  we take the  site variable  
$s_i=0$ for vacancy and $s_i=1,2\dots \nu$ for the different internal states.

As in TASEP, the particles hop towards the right neighbouring site  if it is vacant. However, 
the  hop rate  depends on the internal state  of the particle. Corresponding dynamics is 
\bea
I~ 0 \, \mathop{\longrightarrow}^{\alpha_{I}} \, 0~ I,
\label{eq:dyn1}
\eea
where  $I=1,2,....\nu$ represents the different internal states of the particles and 
$\alpha_I$ are the hop rates of the particles in the $I^{th}$  state.
In addition to  this hopping dynamics, the particles are also allowed to  change their 
internal states when their right neighbouring sites are occupied. In other words
%Thus, the  system evolves  according to  the following set of dynamical rules, 
\bea 
I~ K \, \mathop{\longleftrightarrow}_{p_{JI}}^{p_{IJ}} \, J~ K.\label{eq:dyn2}
\eea
where  the rates $p_{IJ}$ and $p_{JI}$ do not depend on the  state $K$  of the 
neighbouring particle. Note, that $K$ need not necessarily differ from  $I$ or $J$.
This dynamics \eqref{eq:dyn2}  naturally do not show up for $\nu=1$.  The only
dynamics  \eqref{eq:dyn1}   is identical to that of  TASEP.
 
 \begin{figure}[h]
 \includegraphics[width=8cm]{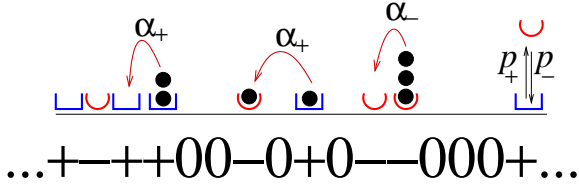}
\caption{Mapping of the exclusion process to the ZRP.  $+$ and $-$ particles are regarded as two 
different kinds of boxes:  shown as square(blue) and semi-circular (red) boxes 
 respectively. Uninterrupted sequence of $0$s in front of $\pm$ are considered as the 
particles in the box.  Particles in this ZRP  can  hop  from  the $\pm$ boxes to the left only,
with rates $\alpha_\pm$.  The empty boxes can change state, $\pm \to \mp$ with rates 
$p_\pm$.}
  \end{figure}

For simplicity we first study the case $\nu=2$ in details;  
generalization to arbitrary $\nu$ is straightforward.
Particles in the two possible internal states here are denoted 
by $+$ and $-$ for notational convenience;  correspondingly the site 
variable $s_i=0, \pm $.
%The  indices $\pm$ of the particles  can be regarded as their (classical) spin states. 
Both $+$ and $-$ particles can move only to the rightward vacant site 
with rates $\alpha_\pm$. Otherwise, if the rightward neighbour is occupied, 
the particles can flip their state $(\pm \to \mp)$ with rates $p_\pm$. 
Explicitly,  
\begin{equation}
\pm 0 \mathop{ \longrightarrow}^{\alpha_{\pm}}  0 \pm ~~~~~~~~~~~
%+ \pm \mathop{\longleftrightharpoons}^{p_{+}}_{p_{-}}  -  \pm ~~~~~~~~~~
+ \pm \mathop{\longleftrightarrow}_{p_{-}}^{p_{+}}  -  \pm ~.
%+ \pm \mathop{\longrightarrow}^{p_{+}}_{p_{-}} -  \pm ~~;~~~~~ - \pm \mathop{\longrightarrow}^{p_{-}} + \pm~~;~~~~~~
\label{eq:model}
\end{equation}

  Clearly, the  total number of particles $N_+ +N_-$ and vacancies $N_0$ are conserved  by this dynamics, 
whereas  individually $N_+$   and $N_-$ are  not. Thus, both the densities $\frac{N_{\pm}}{L}$  
fluctuate and only their averages $\rho_{\pm}= \frac{\la N_{\pm}\ra }{L}$ reach a stationary value  
in the steady state.
%\section{ZRP mapping}

The model can be  mapped to  the Zero  Range Process (ZRP)\cite{tasep,zrp} by identifying $+$ and $-$ as 
two different kinds of boxes denoted by $\tau=\pm$ respectively. Uninterrupted sequence of $0$s, 
say $n$ in number, to the right of  a $+(-)$ particle  is regarded as a $+(-)$ box containing $n$ particles. 
The mapping is described  schematically in Fig. 1.  
Thus, this ZRP with   $N_0$ particles  distributed in 
$M=L-N_0$ boxes  obey the following dynamics: \\
(a) particles from a $+(-)$ box moves to the left box with constant rate $\alpha_+ (\alpha_-)$,
(b) an empty  $+(-)$ box can alter its state  with rate  $p_+ (p_-)$. 

Note that here the hop rates $\alpha_\pm$ of ZRP are independent of  the number of particles in the 
departure box. The box dynamics (b) here can produce correlations among 
boxes (absent in ordinary ZRP) as the non-empty $\pm$ boxes {\it can not} alter their sign.

\section{Steady State for $\nu=2$}

A generic configuration  can be written  as $\{ n_1\tau_1,n_2\tau_2\dots n_M\tau_M\}$ where $n_k$ 
is the number of particles in $k^{th}$ box  of type $\tau_k$. It can be shown that steady state of 
this ZRP has a product measure : 
\beq
P(n_1 \tau_1,n_2\tau_2,\dots n_M \tau_M ) \sim f_{\tau_1}(n_1) f_{\tau_2}(n_2) \dots f_{\tau_M}(n_M)
\label{eq:prodm}
\eeq  
where weight of an individual $\pm$ box   containing  $n$ particles is   $f_{\pm}(n)$. 
\subsection{Proof of Product measure\label{sec:proof} }
   The proof can be constructed using pairwise balance where we find  a unique configuration $C''$ 
for every transition  $C\to C'$ such that 
\beq
P(C) W(C\to C') = P(C'') W(C''\to C).
\label{eq:pb}
\eeq
 It is sufficient to 
consider all possible transitions that  changes the state $n_k \tau_k$ of the box $k$. 
Thus $C=\{\dots  n_{k-1} \tau_{k-1},n_k \tau_k,n_{k+1} \tau_{k+1} \dots\}$. Let us  construct 
 $C''$ for  all possible $C'$s.

\textbf{Case-I $n_k=0$ : } In this case the $k^{th}$ box can change sign. Thus 
$C'=\{\dots  n_{k-1} \tau_{k-1},0\bar \tau_k,n_{k+1} \tau_{k+1} \dots\}$, where 
$\bar\tau_k = -\tau_k$. A choice $C''= C'$ along with  Eq. (\ref{eq:prodm}) and  (\ref{eq:pb})
gives
\beq
\frac{f_-(0)}{f_+(0)}= \frac{ p_+}{ p_-}.
\label{eq:I}
\eeq

{\bf Case II  $n_k >0$ :} In this case a particle from box $k$ can move 
to $k-1$ with rate $\alpha_{\tau_k}$. Thus 
$C'=\{\dots  (n_{k-1}+1) \tau_{k-1},(n_k-1) \tau_k,n_{k+1} \tau_{k+1} \dots\}$. We choose $C''=\{\dots  n_{k-1} \tau_{k-1},(n_k-1) \tau_k,(n_{k+1}+1) \tau_{k+1} \dots\}.$ Then the condition of pairwise balance along with \eqref{eq:prodm} demands
\bea
\alpha_{\tau_k} \frac{f_{\tau_k}(n_k)}{f_{\tau_k}(n_k-1)}=  \alpha_{\tau_{k+1}} \frac{f_{\tau_{k+1}}(n_{k+1}+1)}{f_{\tau_{k+1}} (n_{k+1})}=c \label{eq:II}
\eea
The constant $c$ is independent of $n$ and $\tau$. We set $c=\alpha_+$ without any loss of generality.
Since the steady state weights are  yet to be normalized, we can take $f_+(0)=1$
without loss of generality. Then, Eqs. (\ref{eq:I}) and  (\ref{eq:II}) 
result in 
\beq
f_+(n) = 1; ~~ f_-(n) = p \alpha^n  
\eeq
where  $\alpha= \frac{\alpha_+}{\alpha_-}$  and  $p= \frac{p_+}{p_-}$ are the ratios of 
hop rates and flip rates respectively.

   Now, weights of every configuration in the lattice can be written in terms of the 
steady state weights of ZRP. 
\bea
P(\{s_i\} | i=1,\dots L ) &=& P( \{n_k\tau_k\}| k=1,\dots M ) \cr &=& \prod_{k=1}^M f_{\tau_k}(n_k)
\label{eq:ZRPss}
\eea

     Spatial indices of  the model (\ref{eq:model}) on a lattice are not carried over to 
their corresponding ZRP version,  which makes calculation of the spatial correlations  
unreasonably difficult. We follow a method  discussed in \cite{ozrp}. First, let us 
rewrite the steady state  weights as a matrix product state,  by replacing $s_i$ by 
the corresponding matrices $X_{s_i}$. 
\beq
P(\{s_i\})= Tr[\prod_{i=1}^{L} X_{s_i} ]
\label{eq:MPS}
\eeq
Since we already know
$P(\{s_i\})$ from Eq. \eqref{eq:ZRPss} in terms of  $f_\tau(n)$s, $ X_{s_i}$s must 
be chosen such that 
\beq
Tr[\prod_{i=1}^{L} X_{s_i} ] =\prod_{k=1}^M f_{\tau_k}(n_k).
\label{eq:condition}
\eeq 
 We proceed  with a simpler notation $X_+=D, X_{-}=E$ and $X_0=A$ and 
the following choice,
\bea
D = \ket {d_1}\bra {d_2}; \; E= \ket {e_1}\bra {e_2}.
\label{eq:vec}
\eea
Now, Eq. \eqref{eq:condition} and \eqref{eq:vec} together  impose the 
 following conditions on matrix $A$,  
\bea
 \bra {d_2}  A^n \ket {d_1}  =\bra {d_2} A^n \ket {e_1}= f_+(n)=1\cr
 \bra {e_2}  A^n \ket {e_1} =\bra {e_2}  A^n \ket {d_1} = f_-(n)=p\alpha^n
 \label{eq:matA}
\eea
 Note, that unlike the usual Matrix Product Ansatz\cite{mpaderrida,MPA}, 
 these conditions on  $D$, $E$ and $A$  do not depend on 
the dynamics of the model explicitly. This will be discussed in some length 
later in \ref{sec:MPA}.

An infinite dimensional representation of these  matrices that satisfy
Eq.  \eqref{eq:matA},  similar to those  obtained for the usual ZRP with one kind of boxes \cite{ozrp}, 
can always be constructed. Fortunately for this  model, as $f_\tau(n)$ are simple functions, we have a two 
dimensional representation, 
\bea
 \ket {d_1} = \ket {e_1} =  \ket 1 +\ket 2 ; \bra {d_2} = \bra 1; \bra {e_2} = p \bra 2; \cr
A=  \ket 1 \bra 1 + \alpha \ket 2 \bra 2. 
\eea
Here $\ket 1= \left(\begin{smallmatrix}1 \\ 0\end{smallmatrix}\right)$  and $\ket 2= \left(\begin{smallmatrix}0 \\1\end{smallmatrix}\right) $  are the standard basis vectors in two 
dimensional vector space.  Explicitly, 
\bea
D= \begin{pmatrix} 1 & 0 \\ 1 & 0 \end{pmatrix}; 
 E= \begin{pmatrix} 0 & p \\ 0 & p \end{pmatrix}; A= \begin{pmatrix} 1 & 0 \\ 0 & \alpha \end{pmatrix};\label{eq:dea}
\eea

\subsection{Matrix Product Ansatz \label{sec:MPA}}
   In recent years  steady state weights  of several exclusion models have been obtained applying 
Matrix Product Ansatz (MPA)\cite{mpaderrida,MPA}.  To use MPA   for exclusion models,  first  one writes the steady 
state weight as  the trace of the product of matrices, similar to Eq.  \eqref{eq:MPS}. The matrices 
are required to satisfy  certain algebraic relations, which depends on the dynamics  and also  
involve a set of auxiliary matrices. 

In  this model with $\nu=2$, we already have matrices $D$,  $E$ and $A$ [given by Eq. \eqref{eq:dea}] 
which provide  the  exact steady state weight of the model in matrix product form. However, 
it is not obvious that  these matrices, which are constrained by Eq. \eqref{eq:matA},  also 
follow the algebraic relations  required by MPA. In the following we  show that, in fact for this particular 
dynamics, matrices  of \eqref{eq:dea} can be made to satisfy the required equations with a suitable choice of
the auxiliary matrices.  An explicit representation of  these 
auxiliary matrices $\tilde D$, $\tilde E$ and $\tilde A$ are also given.

The complete set of  equations for the matrices $D$, $E$ and $A$, as required by the MPA, are  
\bea
 \alpha_+ DA &=& -\tilde A D + A \tilde D = \tilde D A - D \tilde A \cr
%- \alpha_- EA = -\tilde E A + E \tilde A ,~~~
 \alpha_- EA &=& -\tilde A E + A \tilde E =\tilde E A - E \tilde A \cr
% - p_+ DD + p_- ED &=& -\tilde D D + D \tilde D &=&  \cr
%- p_+ DE + p_- EE &=& -\tilde D E + D \tilde E \cr
 p_+ DD - p_- ED &=& -\tilde E D + E \tilde D =\tilde D D - D \tilde D \cr
 p_+ DE - p_- EE &=& -\tilde E E + E \tilde E= \tilde D E - D\tilde E,  
%- \alpha_+ DA = -\tilde D A + D \tilde A ,~~~
\eea 

where the auxiliaries  $\tilde D$,$\tilde E$  and  $\tilde A$ also need to be determined.
These set of equations  do not necessarily have a unique solution. For simplicity we  
choose  $\tilde D =D, \tilde E = E$, which  reduces  the above set of equations to,
\bea
p_+ DD = p_- ED&;&\quad p_+ DE = p_- EE\cr 
(1-\alpha_+) DA = D \tilde A &;&\quad
\alpha_+ DA - AD = -\tilde A D\cr
(1-\alpha_-) EA = E \tilde A  &;&\quad
\alpha_- EA - AE = -\tilde A E
\label{eq:mpa}
\eea

Now we need to check, if  $D,E$ and $A$ obtained earlier in Eq. \eqref{eq:dea} satisfy above equation
along with some auxiliary matrix  $\tilde A$. 
 
It turns out that 
$ 
\tilde A = \begin{pmatrix} (1-\alpha_+) & 0 \\ 0 & \alpha(1-\alpha_-) \end{pmatrix}
$ 
consistently  solves the  above equation  along with Eq. \eqref{eq:dea}.

\section{Correlation functions  ($\nu=2$)}

To calculate the partition function  one needs to take care of the conservation of total 
number of  $0$s. Instead, we work in grand canonical ensemble(GCE), where the   
fugacity $z$ associated with `$A$'s fixes the average density of $0$s. Thus the 
 partition function in GCE is 
\beq
Z= Tr[ (D + E  + z A)^L] = Tr[T^L]
\eeq
where we have used $T= D + E  + z A= \begin{pmatrix} 1+z & p \\ 1 & p+z \alpha \end{pmatrix}$ for convenience. The eigenvalues of $T$ are 
\beq
\lambda_\pm = \frac 12 [1 + p + z + \alpha z \pm \sqrt{(1+ p + z + \alpha z)^2 - 4 z(\alpha  + p + \alpha z)} ]
\eeq
Thus, $Z= \lambda_+^L + \lambda_-^L$.
To calculate the correlation functions we need $T^n$.  For large $n$, $\left(\frac{\lambda_-}{\lambda_+}\right)^n \to 0$ and we have  

\begin{figure}[h]
%\vspace*{.1 cm}
 \includegraphics[width=7.5 cm]{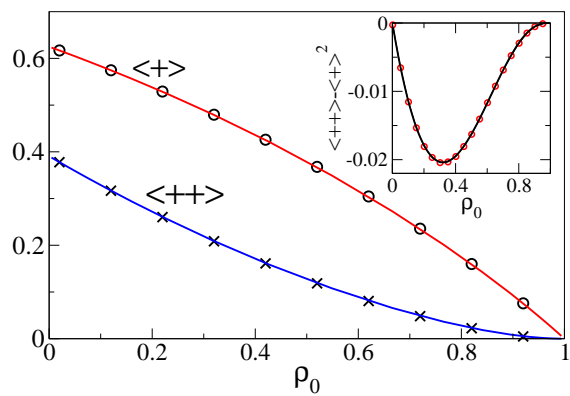}
\caption{ The main figure  compares  Eqs. \eqref{eq:r+}  and  \eqref{eq:r++}  (solid lines) 
with $\la + \ra$ (circle)  and $\la ++\ra$ (cross) obtained from  Monte-carlo simulation  of a
system of  size $L=1000$,   $\alpha =0.3$ and $p=0.6$. The correlation function  $C_{++}$ 
is shown in the inset, where solid line corresponds to  Eq. \eqref{eq:c++}.}
\end{figure}

\bea
T^n_{11} &\simeq&  \frac{\lambda_+ ^n (\lambda_+ -p - \alpha z)}{\lambda_+-\lambda_-}, \qquad
T^n_{12} \simeq  \frac{p\lambda_+ ^n }{\lambda_+-\lambda_-}\cr
T^n_{22} &\simeq& -\frac{\lambda_+ ^n(\lambda_- -p - \alpha z)}{\lambda_+-\lambda_-}, \quad
T^n_{21} \simeq  \frac{\lambda_+ ^n }{\lambda_+-\lambda_-}\n
\eea

\begin{figure}[h]
 \includegraphics[width=7.5cm]{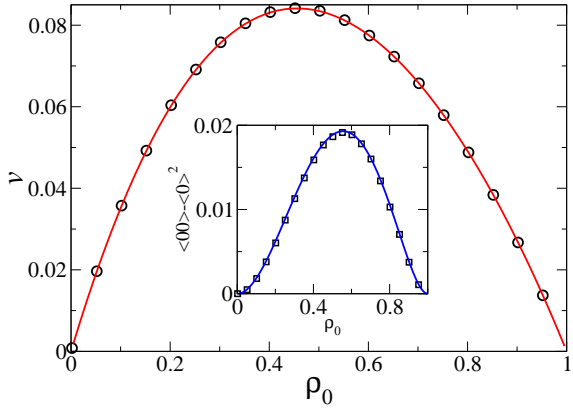}
\caption{ The average velocity of $0$s, obtained from  Eq. \eqref{eq:v}   is 
compared with  the numerical simulation for  $L=1000$,   $\alpha =0.3$ and $p=0.6$. The inset shows   
correlation function  $C_{00}$ versus $\rho_{0}$.}
\end{figure}

Calculation of expectation values of the observables are straightforward in this formulation.  
For example, 
\bea
\rho_0= \frac{\la N_0\ra}{L}&=&  \frac{Tr[zAT^{L-1}]}{Z} = \frac{ z (T^{L-1}_{11} + \alpha T^{L-1}_{22}) }{Z} \label{eq:r0}\\
       &=& \frac{z[(\lambda_+- \alpha \lambda_-)- (1-\alpha)(p+ \alpha z)]}{\lambda_+(\lambda_+-\lambda_-)}\cr
       & & \cr
\rho_+ = \la +\ra & =& \frac{Tr[DT^{L-1}]}{Z} =\frac{ T^{L-1}_{11} + T^{L-1}_{12} }{Z} \cr
                 &=& \frac{(\lambda_+-\alpha z)}{\lambda_+(\lambda_+-\lambda_-)} \label{eq:r+}\\
                 & & \cr
\rho_- = \la - \ra &=& \frac{Tr[ET^{L-1}]}{Z} =\frac{p( T^{L-1}_{21} + T^{L-1}_{22}) }{Z} \cr
             &=& \frac{p(\lambda_+ - z)}{\lambda_+(\lambda_+-\lambda_-)}
\eea 
The average velocity of $0$s is
\bea
v &=& \alpha _-[\alpha  \la +0\ra +  \la -0\ra ]= \alpha_-\frac{z Tr[(\alpha D+E)A T^{L-2}]}{Z} \cr
  &=& \alpha_-\frac{z \alpha (\lambda_+ +p - \alpha z -p (\lambda_- - p - \alpha z))}{\lambda_+ ^2(\lambda_+-\lambda_-)}  \label{eq:v}
\eea

Two point correlations:
\bea
\la ++\ra &=& \frac{Tr[DDT^{L-2}]}{Z}=\frac{ T^{L-2}_{11} + T^{L-2}_{12} }{Z} \cr
         &=& \frac{(\lambda_+-\alpha z)}{\lambda_+ ^2(\lambda_+-\lambda_-)} \label{eq:r++}\\
C_{++} &=& \la ++\ra - \la +\ra ^2 
%    &=& - \frac{(\lambda_+-\alpha z)(\lambda_--\alpha z)}{\lambda_+ ^2(\lambda_+-\lambda_-)^2} \cr
  = - \frac{ p z(1-\alpha)}{\lambda_+ ^2(\lambda_+-\lambda_-)^2}  \label{eq:c++}\\
& & \cr
\la 00\ra &=& \frac{z^2 Tr[AAT^{L-2}]}{Z}=\frac{z^2 (T^{L-2}_{11} + \alpha^2 T^{L-2}_{12}) }{Z} \cr
 &=& \frac{z^2[(\lambda_+- \alpha^2 \lambda_-)- (1-\alpha^2)(p+ \alpha z)]}{\lambda_+^2(\lambda_+-\lambda_-)}  \label{eq:c00}\\
& & \cr
C_{00} &=&\la 00\ra - \la 0\ra ^2 
      = \frac{ p z^2(1-\alpha)^2}{\lambda_+ ^2(\lambda_+-\lambda_-)^2}
\eea

 We have compared these results with  those obtained from the Monte-carlo 
simulations.  In Fig. 2 we have  shown variation of $\la + \ra$ and $\la ++\ra$ with 
$\rho_0$. The parameters $\alpha =0.3$, $p=0.6$ and  the system size $L=1000$ were fixed. 
Solid lines correspond  to Eqs. \eqref{eq:r+}  and  \eqref{eq:r++}  respectively, where 
$\rho_0$ was obtained from  Eq.  \eqref{eq:r0}. 
Inset of Fig. 2 compares Eq. \eqref{eq:c++} with $C_{++}$ obtained  from simulations.     

Similarly, in  Fig. 3 we have  compared the average velocity of $0$s with  Eq. \eqref{eq:v}. The inset 
therein shows  variation of $C_{00} = \la 00\ra -\la0\ra^2$.

\section{General results for $\nu>2$}
  In this section we will extend the  results of section \ref{sec:proof} and obtain the exact 
steady state weights of the model with generic $\nu>2$. 
  
Here, it is evident that  the dynamics \eqref{eq:dyn1} and \eqref{eq:dyn2} can also be mapped to 
that of a ZRP with $\nu$ kinds of boxes, where a particle from a randomly chosen box of kind $I$ moves to the left box with a 
constant rate $\alpha_I$ and an empty box can change its internal state.
As before  we look for a factorized steady state  satisfying pairwise balance condition. Accordingly,
for non-empty boxes, we  obtain 
%again obtain Eq. \eqref{eq:II} which results in
\bea 
\alpha_I \frac{ f_I(n+1)}{f_I(n)} = c, \forall I=1,2\dots \nu
\eea
similar to  Eq. \eqref{eq:II}. Here again $c$ is an arbitrary constant, independent of the number
of particles $n$ and the state of the box $I$, can be taken to be $1$ without any loss of generality.
Thus $f_I(n) = f_I(0)/ \alpha_I^n$. The relative weights of the empty boxes, again, satisfy
\bea 
 {f_I(0) \over f_J(0)} &=&{ p_{JI} \over p_{IJ} } \label{eq:pij}
\eea
for all possible pairs $(I,J)$. This condition, which is a generalized version of the Eq. \eqref{eq:I}, 
demands that the steady state cannot have  a product measure form if all the  $\nu(\nu-1)$  rates  $p_{IJ}$s 
are independent; they must be related  in the following way %set of equations %$ (\nu-1)(\nu-2)/2$ constraints

\bea 
{p_{IJ}\over p_{JI}}{p_{JK}\over p_{KJ}}={p_{IK}\over p_{KI}} \quad \forall I,J,K \label{eq:pijk}
\eea

This set of  equations imposes  $(\nu-1)(\nu-2)/2$ constraints leaving  $(\nu-1)(\nu/2+1)$  
independent  $p_{IJ}$s.  
Further, we take $f_1(0)=1$ as the weights are not  normalized yet, which results in 
\bea
f_1(n)= {1 \over \alpha_1^n} \;\quad {\rm and}\;\quad f_I(n)=  {p_{1I}\over p_{I1}}{1 \over \alpha_I^n} \quad \forall I>1. 
\eea

%We follow the similar procedure as in the two species case to

To find the spatial correlation functions we rewrite the steady state in a matrix product 
form which is a direct generalization of the two species case; 
each particle of species $I$ is replaced by $D_I$ and vacancies by $A$.  
The non-commuting  set of matrices $D_I$ and $A$ must satify Eq. (11) with the above $f_I$.
This can be achieved by  choosing $D_I= \ket d \bra {d_I}$, similar to Eq. \eqref{eq:vec}, 
resulting in  
\bea
\bra {d_I} A^n \ket {d} = f_I(n)\n
\eea
 which is a generalized form of Eq. \eqref{eq:matA}.

It is straightforward to find a $\nu$ dimensional representation of these matrices; 
$\ket d = \sum_i \ket i$, $\bra {d_I}= {p_{1I}\over p_{I1}} \bra i$ and $A=Diag ({1 \over \alpha_1}, \dots {1 \over \alpha_\nu}).$ 
Here, $\{\ket i\}$ are the standard basis set for the $\nu$-dimensional vector space.

Let us discuss the $\nu=3$ case in some details.  Explicitly, the dynamics is 
\bea
10  \mathop{\longrightarrow}^{\alpha_1} 01;\;& 20  \mathop{\longrightarrow}^{\alpha_2} 02; \; &
30  \mathop{\longrightarrow}^{\alpha_3} 03 \cr 
1I  \mathop{\longleftrightarrow}_{p_{21}}^{p_{12}} 2I;\;& 2I \mathop{\longleftrightarrow}_{p_{32}}^{p_{23}} 3I; \;&
1I \mathop{\longleftrightarrow}_{p_{31}}^{p_{13}} 3I,\;\; \forall I\ne 0.\n 
\eea
The model is mapped to a ZRP with three different kinds  of boxes. Its  steady state has product measure  
only when the rates $p_{IJ}$ follow   Eq. \eqref{eq:pijk}; 
 $${p_{12}\over p_{21}} {p_{23}\over p_{32}}={p_{13}\over p_{31}}.$$ 
%there is no restriction on $\alpha$ 
Thus, any five out of these six $p_{IJ}$s  can be chosen indepedently and the sixth one  is 
fixed by the above equation.

 The spatial correlations of this system on the lattice  can be calculated  in a straightforward manner 
using the prescription decribed  for $\nu=2$.  All one needs is  the  explicit representation of  the 
matrices, which are,
\bea
D_1 = \begin{pmatrix} 1 & 0 & 0 \\  1 & 0 & 0 \\  1 & 0 & 0 \end{pmatrix};~ D_2 = {p_{12} \over p_{21}}\begin{pmatrix} 0 & 1 & 0 \\  0 & 1 & 0 \\  0 & 1 & 0 \end{pmatrix}; \cr
D_3 = {p_{13} \over p_{31}}\begin{pmatrix} 0 & 0 & 1 \\  0 & 0 & 1 \\ 0 & 0 & 1 \end{pmatrix};~ A = \begin{pmatrix} {1 \over \alpha_1} & 0 & 0 \\  0 & {1 \over \alpha_2} & 0 \\  0 & 0 & {1 \over \alpha_3} \end{pmatrix};
\eea

The correlation functions here are found to be  qualitatively same as that  of 
$\nu=2$ case.

\section{Discussion and Conclusion }

  Several  multiple species exclusion processes   with or without conservation  have  been studied earlier, 
mostly numerically, in different contexts. In this article we study a  multispecies exclusion model 
where  exact  steady state weights and spacial correlations are  calculated   analytically. 
  
\begin{figure}[h]
\vspace*{0.5cm}
 \includegraphics[width=8 cm]{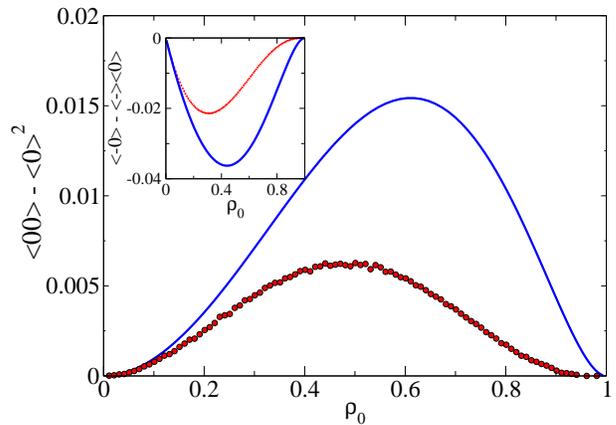}
\caption{ Spatial correlations  of  the conserved  model  obtained from  Monte-Carlo simulations (symbols) on 
a system of size $L=1000$ are compared  with  exact correlation functions of  the  
non-conserved  model (line) with  $p=0.8$ \cite{note1}.  For both the models 
$\alpha=0.5$.}\label{fig:compare}
\end{figure}

This model with $\nu=1$  is identical to TASEP, which,  on a ring,  has  spatially  uncorrelated steady state.
When  internal degrees of freedom are introduced ($i. ~e.~ \nu>1$), two new  features  appear in the dynamics 
: i) particles in different internal states hop with different rates, ii) the particles can change their 
internal states resulting  in non-conservation of number of particles belonging to  each state. 
To investigate  whcih of these features generates spatial correlations we revisit $\nu=2$ case of the model 
with  the nonconserving  part of the dynamics \eqref{eq:dyn2}  replaced by  a conserving one,  
\bea
\pm 0 \mathop{ \longrightarrow}^{\alpha_{\pm}}  0 \pm  ~~~~~~~~~~~
~ +- \mathop{ \longleftrightarrow}^{1}_{1} -+ ~.
\label{eq:new} 
\eea
Symmetic exchange of   $+$ and $-$  particles  ensures that particle current  gets contribution only from 
the hoping dynamics,  as  was the case in the non-conserved syetm.

The  most generic particle conserving  two species model $\tau \tau^\prime\to \tau^\prime \tau$  with  rate $w_{\tau\tau^\prime}$, 
where $\tau \ne \tau^\prime \in 0,1,2$ 
has been discussed in Section 7 of  the review article \cite{MPA}.  It has been shown there that the steady state of this 
generic model can be obtained using Matrix Product Ansatz, when the rates satisfy a set of conditions, 
Eqs.(7.8)- (7.13) therein.  The particle conserving model \eqref{eq:new} discussed here is a special case of the above 
with  $w_{10}=\alpha_+,w_{20}=\alpha_-,w_{12} =1=w_{21}$, and  all other rates are zero.
%of Ref.  \cite{MPA}. 
Here, MPA   provides an exact solution only when $\alpha_+ = \alpha _-$;  corresponding 
steady state  turns out to be spatially uncorrelated. This indicates that  
{\it non-conservation} is  irrelevant, at least for  $\alpha_+ = \alpha _-$, as 
the   non-conserving  model \eqref{eq:model} too provides  an uncorrelated steady state in this case. 
To understand, if non-conservation plays any role in generating spatial correlations, we  
study  model \eqref{eq:new}   numerically for generic  rates  $\alpha_\pm$. These 
Monte-Carlo simulations  reveal that non-trivial spatial correlations 
appear  for $\alpha_+ \ne \alpha _-$. 
In Fig. \ref{fig:compare}  we compare  some of the two-point correlations  obtained  
from Monte-Carlo simulations  of the conserved system with the known exact results of  
the corresponding \cite{note1} non-conserved model.  Clearly the spatial correlations  of the 
conserved model  are  qualitatively similar to those of the  non-conserved 
case.  Thus, conservation of internal degrees  is an irrelevant criterion in developing
spatial correlations in exclusion processes, rather the unequal hop rates are responsible.   
  
In conclusion, we  have studied  an  exclusion process on a one dimensional system with periodic 
boundary, where particles carry $\nu$ internal degrees of freedom.  Along with a directional  hopping dynamics that 
depends on the internal state of  the particle   non-conservation   is introduced  by  allowing   
a particle to change its  internal state when the target site is occupied.  
We show that this model  can be mapped to a zero range process  where  particles  distributed 
in  $\nu$ diffrent kinds of   boxes hop with  different rates and empty boxes can change their state.
 The steady state weights  of the exclusion process could be written  in matrix product form using the 
exact single box weights obtained from ZRP. Exact spatial correlations have been calculated  exploiting 
the  matrix product form.  From the comparison of these correlations  with  those of a
corresponding conserved  model we conclude  that the unequal hop rates of particles belonging to 
different internal states  is the possible cause  for the non-trivial spatial correlations 
observed on a ring.

{\it Acknowledgements :}  The authors would like to thank M. Barma and H. Sachdev for discussing 
their ongoing work on a  similar two-species model.    UB would like to acknowledge thankfully 
the financial support of the Council of Scientific and Industrial Research, India (SPM-07/489(0034)/2007).


\begin{thebibliography}{99}
\bibitem{note1} The density of  particles  in  any internal state  in  the conserved model, which is fixed, 
corresponds to the  average density of particles in the same internal state for the  non-conserved model.
\end{thebibliography}
\end{document}